\documentstyle[12pt]{article}\textwidth 16.5cm
\begin{document}
%
%
\begin{titlepage}
\begin{raggedleft}
THES-TP 95/05\\
June 1995\\
\end{raggedleft}
\vspace{2em}
\begin{center}
{\Large\bf{ Hamiltonian reduction of the $U_{EM}(1)$\\
\vspace{0.25em}
 gauged three flavour WZW model}}\\
\vspace{2em}
{\large J.E.Paschalis and P.I.Porfyriadis}
\footnote{On leave from the Tbilisi State University, Tbilisi,
Georgia.}
\vspace{1em}\\
Department of Theoretical Physics, University of
Thessaloniki,\\GR 54006 \hspace{0.25em},\hspace{0.25em} Thessaloniki
\hspace{0.25em},\hspace{0.25em} Greece\\
\vspace{1em}
{\scriptsize PASCHALIS@OLYMP.CCF.AUTH.GR\\
PORFYRIADIS@OLYMP.CCF.AUTH.GR}
\end{center}
\vspace{2em}
\begin{abstract}
The three-flavour Wess-Zumino model coupled to electromagnetism
is treated as a constraint system using the Faddeev-Jackiw
method. Expanding into series of powers of the Goldstone
boson fields and keeping terms up to second and third order
we obtain Coulomb-gauge hamiltonian densities.
\end{abstract}
\end{titlepage}
%
%
\section{Introduction}
\label{intro}

The Faddeev-Jackiv method \cite{F-J}
provides a simple and straightforward way to deal with
constraint systems without having to distinguish between primary
and secondary, first class and second class constraints.
 A lot of work has been done in this
field by several authors. The equivalence of the Faddeev-Jackiw
approach to the Dirac's method is discussed in
\cite{Govaerts,Montani}. Its extension in supersymmetry is given
in \cite{Kulsh,Barcelos}. Application of the approach to the
light cone quantum field theory is given in \cite{Jun,Jacob},
and to hidden symmetries in \cite{Wotzasek}. Further elaboration
and applications of the method in different cases is given in
\cite{Bar-Wot} . A formulation, intimately related to the FJ
approach, for constructing unconstraint hamiltonians starting
from general first order lagrangians is given in
\cite{Cronstrom}.

In a previous paper \cite{PP} we used the Faddeev-Jackiw method
\cite{F-J} to treat the two-flavour WZW model coupled
to electromagnetism \cite{Witten,Donoghue}. In this work we
extend this analysis to the SU(3) case.
 The gauge invariant action under the
electromagnetic group gauge transformations is given by
\begin{eqnarray}
  \Gamma_{eff}(U,A_\mu)&\!\!\!\! =&\!\!\!\! {\Gamma}_{EM} (A_\mu) +
                 \Gamma_\sigma (U,A_\mu) + {\Gamma}_{WZW} (U) +
                       {\Gamma}_{WZW} (U,A_\mu)\; \; ,
\end{eqnarray}
where the first term is the action of the free electromagnetic
field and the sum of the rest three terms constitute the action
of the gauged Wess-Zumino-Witten model
\begin{eqnarray}
\nonumber \\
  {\Gamma}_{EM} (A_\mu)&\!\!\!\!=&\!\!\!\! -\frac{1}{4}\int\!d^4\!x
                 F_{\mu \nu} F^{\mu \nu}\; \; ,\nonumber \\
\nonumber \\
  \Gamma_\sigma (U,A_\mu)&\!\!\!\! =&\!\!\!\! -\frac{f_\pi^2}{16}
                 \int\!d^4\!x \mbox{tr}\,(R_\mu R^\mu)\nonumber \\
                 &\!\!\!\!=&\!\!\!\!-\frac{f_\pi^{2}}{16}
                 \int\!d^4\!x \mbox{tr}\,(r_\mu r^\mu) +
                 \frac{if_\pi^{2}e}{8}
                 \int\!d^4\!x A_\mu \mbox{tr}\,[Q(r_\mu-l_\mu)]
                 \nonumber \\ &\!\!\!\!+&\!\!\!\! \frac{f_\pi^{2}e^2}{8}
                 \int\!d^4\!x A_\mu A^\mu \mbox{tr}\,
                 (Q^2-U^\dagger QUQ)\; \; , \\
\nonumber \\
   {\Gamma}_{WZW}(U)&\!\!\!\!=&\!\!\!\!-\frac{iN_c}{240\pi^2}
            \int\!d^5\!x\epsilon^{ijklm}
        \mbox{tr}\,(l_{i}l_{j}l_{k}l_{l}l_{m})\; \; ,\nonumber \\
\nonumber \\
  {\Gamma}_{WZW} (U,A_\mu)&\!\!\!\! =&\!\!\!\! -\frac{N_{c}e}{48\pi^2}
                 \int\!d^4\!x \epsilon^{\mu \nu \alpha \beta}
                  A_\mu \mbox{tr}\,[Q(r_\nu r_\alpha r_\beta +
                  l_\nu l_\alpha l_\beta)]\nonumber \\
                  &\!\!\!\!+&\!\!\!\! \frac{iN_{c}e^2}{24\pi^2}
                  \int\!d^4\!x \epsilon^{\mu \nu \alpha \beta}
                  A_\mu(\partial_\nu A_\alpha)
                  \mbox{tr}\,[Q^2(r_\beta+l_\beta)
                  + \frac{1}{2}QU^\dagger QUr_\beta
                  +\frac{1}{2}QUQU^\dagger l_\beta] \nonumber \; \; ,\\
\nonumber
\end{eqnarray}
where $\;\; U=\exp\,(2i\theta_a\lambda_a/f_\pi)\;\;,\;\; a=1,...,8\;\;$
 is an element of SU(3)
\begin{eqnarray}
   r_\mu=U^\dagger\partial_\mu U\; \; ,\hspace{1em}
   R_\mu=U^\dagger D_\mu U\; \; ,\hspace{1em}
   l_\mu=(\partial_\mu U)U^\dagger \; \; ,\hspace{1em}
   L_\mu=(D_\mu U)U^\dagger\; \; .\nonumber
\end{eqnarray}
See Appendix for notation.

The model describes succesfully the low energy dynamics of
photon and the eight Goldstone bosons including their
interactions related to the axial anomaly.

As in the two-flavour case we expand the U fields into series of
powers of the Goldstone boson fields $\;\; \theta_{a}$ .
Then using
the expression for $\Gamma_{eff}(U,A_\mu)$ given in (1)
we obtain lagrangian
densities containing terms initially up to two and afterwards
up to three Goldstone boson fields. These lagrangian densities
are constrained and we treat them
using the Faddeev-Jackiw approach. The general
case which involves any number of Goldstone bosons is still out
of hand.
\section{Expanding up to second order in the Goldstone boson fields}
\label{exp-sec}

We expand the U field into series of powers of the Goldstone
boson fields
\begin{displaymath}
   U=1+\frac{2i}{f_\pi}\theta_{a}\lambda_{a}+...\;\;,\;\;\;
     \;\; a=1,...,8
\end{displaymath}
and we substitute back into the expression (1) for the effective
action. The resulting lagrangian density with terms up to second
order in $\theta_{a}$ is given by
\begin{eqnarray}
  {\cal L}_{eff}&\!\!\!\!=&\!\!\!\!{\cal L}_{EM}+
                 {\cal L}_{\sigma}^{(2)}+{\cal L}_{WZW}^{(2)}
                 +O(\theta^3)\; \; ,\\
  {\cal L}_{EM}&\!\!\!\!=&\!\!\!\!-\frac{1}{4}
                F_{\mu \nu}F^{\mu \nu}\; \; ,\nonumber \\
  {\cal L}_{\sigma}^{(2)}&\!\!\!\!=&\!\!\!\! \frac{1}{2}
               (\partial_\mu \theta_{a})(\partial^\mu\theta_{a})+
                eA^\mu(\theta_{2}\partial_\mu\theta_{1}-
                \theta_{1}\partial_\mu\theta_{2}+
                \theta_{5}\partial_\mu\theta_{4}-
                \theta_{4}\partial_\mu\theta_{5}) \nonumber \\
      &\!\!\!\!+&\!\!\!\! \frac{e^2}{2}A_\mu A^\mu
          (\theta_{1}^{2}+\theta_{2}^{2}+\theta_{4}^{2}+\theta_{5}^{2})
                        \; \; ,\nonumber \\
  {\cal L}_{WZW}^{(2)}&\!\!\!\!=&\!\!\!\!-\frac{N_{c}e^2}{12\pi^{2}f_\pi}
               {\epsilon^{\mu \nu \alpha \beta}}A_\mu(\partial_\nu
               A_\alpha)(\partial_\beta \theta_{3}+\frac{1}{\sqrt{3}}
                         \partial_\beta\theta_{8}) \; \; .\nonumber
\end{eqnarray}

We can easily check that the $U_{EM}(1)$ gauge invariance is not lost.
In the two flavour case \cite{PP} the expression for the corresponding
lagrangian density can be deduced from (3) by omitting those
terms which include powers or derivatives of $\theta_4$ up to
$\theta_8$. We should also notice that the term
$\Gamma_{WZW}(U)$ of the effective action given in (2) which is
zero in the two flavour case, contributes, in the three flavour
case, terms involving five or more Goldstone boson fields.
 In the non-covariant notation, (3) can be written as
follows
\begin{eqnarray}
   {\cal L}_{eff}&\!\!\!\!=&\!\!\!\!-\bf E \cdot \dot{\bf A}\rm+
               \frac{1}{2}\dot{\theta_{\it a}}^{2}+
               eA_{0}(\theta_{2}\dot{\theta_{1}} -
                      \theta_{1}\dot{\theta_{2}}+
                      \theta_{5}\dot{\theta_{4}} -
                      \theta_{4}\dot{\theta_{5}})+
               A_{0}\bf \nabla \cdot E\rm\nonumber \\
       &\!\!\!\!+&\!\!\!\! \frac{e^2}{2}(A_{0}^{2}-{\bf A}^{2}\rm)
               (\theta_{1}^{2}+ \theta_{2}^{2}+
                \theta_{4}^{2}+ \theta_{5}^{2})-
               \frac{1}{2}(\bf E^{\rm 2}+ B^{\rm 2} +
               (\nabla\rm\theta_{\it a})^{2})\nonumber \\
       &\!\!\!\!+&\!\!\!\! e\bf A\cdot\rm(\theta_{2}\bf\nabla
                \rm\theta_{1}-\theta_{1} \bf \nabla\rm \theta_{2}+
                \theta_{5}\bf\nabla\rm\theta_{4}-
                    \theta_{4} \bf\nabla\rm \theta_{5})
                 + \frac{N_{c}e^2}{6\pi^{2}f_\pi}
                 (\bf E \cdot B)\rm (\theta_{3}+
                         \frac{1}{\sqrt{3}}\theta_{8})\; \; ,
\end{eqnarray}
where $\bf E=-\dot{\bf A}-{\nabla}\rm A_{0},\hspace{1em}
\bf B=\bf{\nabla \times A}\; \; $ are the electric and magnetic
fields. We see that the dynamical variables in (4) which span the
configuration space of the lagrangian density are the three
components of the vector potential and the eight Goldstone boson
fields. The scalar potential $A_{0}$ will be treated not as a
dynamical variable but as a Lagrange multiplier since there is
no time derivative of it in (4). Now according to Faddeev and
Jackiw \cite{F-J} we must construct out of (4),
which is second order in
time derivatives of the $\theta_{a}$ fields, a new lagrangian
density which is first order in time derivatives of all the
dynamical variables. This can be accomplished by enlarging the
configuration space so that it includes the whole phase space of
the model. The enlarged configuration space coordinates will
include the three vector potential components, the eight
$\theta_{a}$ fields and the corresponding canonical momenta.

The canonical momenta conjugate to $\bf A $ are the three
component of $- \bf E $, and those conjugate to $\theta_{a}$
are given by
\begin{displaymath}
 \begin{array}{lllll}
\vspace{0.5em}
  p_{1}=\frac{\partial {\cal L}_{eff}}{\partial \dot{\theta}_1}=
        \dot{\theta}_{1}+eA_{0}\theta_{2}\;\;,\\
\vspace{0.5em}
  p_{2}=\frac{\partial {\cal L}_{eff}}{\partial \dot{\theta}_2}=
         \dot{\theta}_{2}-eA_{0}\theta_{1}\;\;,\\
\vspace{0.5em}
  p_{3}=\frac{\partial {\cal L}_{eff}}{\partial \dot{\theta}_3}=
        \dot{\theta}_{3}\; \; ,\\
\vspace{0.5em}
  p_{4}=\frac{\partial {\cal L}_{eff}}{\partial \dot{\theta}_4}=
        \dot{\theta}_{4}+eA_{0}\theta_{5}\;\;,\\
\vspace{0.5em}
  p_{5}=\frac{\partial {\cal L}_{eff}}{\partial \dot{\theta}_5}=
        \dot{\theta}_{5}-eA_{0}\theta_{4}\;\;,\\
\vspace{0.5em}
  p_{6}=\frac{\partial {\cal L}_{eff}}{\partial \dot{\theta}_6}=
        \dot{\theta}_{6}\; \; ,\\
\vspace{0.5em}
  p_{7}=\frac{\partial {\cal L}_{eff}}{\partial \dot{\theta}_7}=
        \dot{\theta}_{7}\; \; ,\\
\vspace{0.5em}
  p_{8}=\frac{\partial {\cal L}_{eff}}{\partial \dot{\theta}_8}=
        \dot{\theta}_{8}\; \; .
 \end{array}
\end{displaymath}
So the resulting, first order in time derivatives, lagrangian
density is given by
\vspace{1em}
\begin{eqnarray}
  {\cal L}_{eff}&\!\!\!\!=&\!\!\!\!-\bf E\cdot\dot{A}\rm+
                 p_{\it a}\dot{\theta_{\it a}}-
                 H^{(2)}(\bf E,A,\rm p_{\it a},\theta_{\it a})
                -A_{0}(\rho^{(2)}-\bf\nabla\cdot E\rm)
                +O(\theta^3) \;\; , \\
\nonumber \\
 H^{(2)} &\!\!\!\!=&\!\!\!\!\frac{1}{2}[\bf E^{\rm 2}+\bf B^{\rm 2}
              +(\nabla \rm\theta_{\it a})^{2}+p_{\it a}^{2}]
              +\frac{e^2}{2}\bf A^{\rm 2}\rm
          (\theta_{1}^{2}+\theta_{2}^{2}+\theta_{4}^{2}+\theta_{5}^{2})
                 \nonumber \\
         &\!\!\!\!+&\!\!\!\!
              e\bf A\cdot\rm(\theta_{1}\bf \nabla \rm \theta_{2}-
                              \theta_2 \bf \nabla \rm \theta_1+
                              \theta_{4}\bf \nabla \rm \theta_{5}-
              \theta_5 \bf \nabla \rm \theta_4) -
              \frac{N_{c}e^2}{6\pi^2 f_\pi}\bf
   (E\cdot B)\rm(\theta_{3}+\frac{1}{\sqrt{3}}\theta_{8})
                     \; ,\nonumber \\
\nonumber \\
\rho^{(2)}&\!\!\!\!=&\!\!\!\! e(p_{2}\theta_1-p_{1}\theta_{2}+
               p_{5}\theta_4-p_{4}\theta_{5}) \; \; .\nonumber
\end{eqnarray}
\vspace{0.5em}

The expression $ (p_{a}\dot{\theta_{a}}-\bf E \cdot \dot{A}\rm)dt$,
called the canonical one-form, can be written up to a total time
derivative as
\begin{eqnarray}
   \frac{1}{2}\xi^i \omega_{ij} d\xi^j \;\; ,
\end{eqnarray}
where
\begin{eqnarray}
  \xi^i &\!\!\!\!=&\!\!\!\! A_i \;\; ,\;\; i=1,...,3 \nonumber \\
  \xi^i &\!\!\!\!=&\!\!\!\! \theta_i \;\; ,\;\; i=4,...,11 \nonumber\\
  \xi^i &\!\!\!\!=&\!\!\!\! -E_i \;\; ,\;\; i=12,...,14 \nonumber\\
  \xi^i &\!\!\!\!=&\!\!\!\! p_i \;\; ,\;\; i=15,...,22 \nonumber
\end{eqnarray}
and $\omega_{ij}$ is the symplectic $ 22 \times 22 $ matrix
\begin{displaymath}
 \omega_{ij}=\left( \matrix{0 & I \cr -I& 0 \cr} \right)_{ij}
\end{displaymath}
As we mentioned before $A_{0}$ is the Lagrange multiplier.
The expression $ \rho^{(2)}- \nabla\cdot\bf E\rm $
which is multiplied by $A_{0}$
is the only constraint in (5). The next step is to solve the
equation of the constraint
\begin{eqnarray}
  \nabla\cdot\bf E \rm -\rho^{(2)}=0 \;\; ,
\end{eqnarray}
and incorporate the solution into the expression (5) for the
lagrangian density. In order to do that we decompose the
electric field $\bf E$
and the vector potential $\bf A$ into transverse and
longitudinal components
\begin{displaymath}
  \bf E=E_{\rm T}+E_{\rm L}\;\; , \;\; \hspace{1em}
      A=A_{\rm T}+A_{\rm L}\;\; , \;\;
\end{displaymath}
\begin{displaymath}
   \bf \nabla \cdot E_{\rm T}=\rm 0 \;\; , \;\;
   \bf \nabla \times E_{\rm L}=\rm 0 \;\; , \;\;
   \bf \nabla \cdot A_{\rm T}=\rm 0 \;\; , \;\;
   \bf \nabla \times A_{\rm L}=\rm 0\; \; .
\end{displaymath}
Then (7) implies that
\begin{equation}
\bf E_{\rm L}\rm=\frac{\bf \nabla \rm}{\bf \nabla
\rm^2}\rho^{(2)}\equiv-\frac{1}{4\pi}\int\!d^3\!y
\frac{1}{\mid \bf y \rm -\bf x \rm \mid}\nabla \rho^{(2)}(y) \; \; .
\end{equation}
Substituting into (5) the expression of $\bf E_{\rm L}$ given in
(8) we get (apart from a total divergence)
\vspace{0.10em}
\begin{eqnarray}
  {\cal L}_{eff} &\!\!\!\!=&\!\!\!\!
                     -\bf E_{\rm T}\cdot\dot{A}_{\rm T}\rm
                     +\rho^{(2)}\frac{\bf \nabla \rm}
                     {\bf \nabla \rm^2}\cdot\dot{\bf A}_{\rm L}\rm
                      +p_{\it a}\dot{\theta}_{\it a} - \frac{1}{2}
                     [\bf E_{\rm T}^{\rm 2}+ B^{\rm 2}-
                     \rm \rho^{(2)}\frac{1}{\bf \nabla \rm^2}
                     \rho^{(2)}+(\nabla \rm\theta_{\it a})^{2}+
                     p_{\it a}^{2}] \nonumber \\
     &\!\!\!\!-&\!\!\!\! e\bf A\cdot\rm
                (\theta_{1} \bf \nabla \rm \theta_{2}-
                \theta_2 \bf \nabla \rm \theta_1 +
                \theta_{4} \bf \nabla \rm \theta_{5}-
                 \theta_5 \bf \nabla \rm \theta_4) -
                 \frac{e^2}{2}\bf A\rm^2
                 (\theta_{1}^{2}+\theta_{2}^{2}+
                  \theta_{4}^{2}+\theta_{5}^{2})\nonumber\\
      &\!\!\!\!+&\!\!\!\!
         \frac{N_{c}e^2}{6\pi^2 f_\pi}[\bf E_{\rm T}\cdot B\rm +
         (\frac{\bf \nabla \rm}{\bf \nabla \rm^2}\rm \rho^{(2)})
         \bf\cdot B\rm] (\theta_3+\frac{1}{\sqrt{3}}\theta_8)\; \; .
\end{eqnarray}
\vspace{0.25em}
We see that the canonical one-form in (9) has lost the standard
form as given in (6) because of the term
$\rho^{(2)}\frac{\bf \nabla \rm}
{\bf \nabla \rm^2}\cdot\dot{\bf A}_{\rm L}\rm dt $. According
 to Darboux's theorem \cite{F-J} we can perform  coordinate
transformations so that the canonical one-form gets back the
standard form. This coordinate transformations called Darboux's
transformations can be written as follows
\vspace{1em}
\begin{displaymath}
   p_1\rightarrow p_1 \cos{\alpha}+p_2\sin{\alpha}
           \;\; , \;\; \hspace{1em}
   \theta_1 \rightarrow \theta_1 \cos{\alpha}+\theta_2 \sin{\alpha}
            \;\; , \;\;
\end{displaymath}
\begin{displaymath}
   p_2 \rightarrow p_2 \cos{\alpha}-p_1\sin{\alpha}
           \;\; , \;\; \hspace{1em}
   \theta_2 \rightarrow \theta_2 \cos{\alpha}-\theta_1 \sin{\alpha}
           \;\; , \;\;
\end{displaymath}
\begin{equation}
   p_4\rightarrow p_4 \cos{\alpha}+p_5\sin{\alpha}
           \;\; , \;\; \hspace{1em}
   \theta_4 \rightarrow \theta_4 \cos{\alpha}+\theta_5 \sin{\alpha}
           \;\; , \;\;
\end{equation}
\begin{displaymath}
   p_5 \rightarrow p_5 \cos{\alpha}-p_4\sin{\alpha}
           \;\; , \;\; \hspace{1em}
   \theta_5 \rightarrow \theta_5 \cos{\alpha}-\theta_4 \sin{\alpha}
           \;\; , \;\;
\end{displaymath}
\begin{displaymath}
   p_3 \rightarrow p_3 \;\; , \;\; \theta_3 \rightarrow
\theta_3 \;\; , \;\;
   p_6 \rightarrow p_6 \;\; , \;\; \theta_6 \rightarrow
\theta_6 \;\; , \;\;
   p_7 \rightarrow p_7 \;\; , \;\; \theta_7 \rightarrow
\theta_7 \;\; , \;\;
\end{displaymath}
\begin{displaymath}
   p_8 \rightarrow p_8 \;\; , \;\; \theta_8 \rightarrow
\theta_8 \;\; , \;\;
  \bf E_{\rm T} \rightarrow E_{\rm T}\;\; , \;\; A_{\rm T}
        \rightarrow A_{\rm T}\;\; , \;\;
\end{displaymath}
where
$\alpha = e\frac{\bf \nabla \rm}{\bf \nabla \rm^2}\cdot\bf A_{\rm L}$.
 We substitute (10) into the expression (9) for the lagrangian
density and we find out that all the terms which include powers or
derivatives of $\bf A_{\rm L}$ cancel out exactly. Only terms
with the physical transverse components of the vector potential
survive, and the canonical one-form in (9) acquires the standard
form. The resulting lagrangian density is given by
\begin{eqnarray}
  {\cal L}_{eff}=-\bf E_{\rm T}\cdot\dot{A}_{\rm T}+
          \rm p_{\it a}\dot{\theta_{\it a}} - H_{C}^{(2)} \; \; ,
\end{eqnarray}
where
\begin{eqnarray}
   H_{C}^{(2)}&\!\!\!\!=&\!\!\!\!\frac{1}{2}
        [\bf E_{\rm T}^{\rm 2}+B^{\rm 2}\rm-
                  \rho^{(2)}\frac{1}{\bf \nabla \rm^2}
                  \rho^{(2)}+(\bf \nabla \rm \rm\theta_{\it a})^{2}
                  +p_{\it a}^{2}]
          + e\bf A_{\rm T}\cdot\rm
                 (\theta_{1} \bf \nabla \rm \theta_{2}-
                  \theta_2 \bf \nabla \rm \theta_1+
                  \theta_{4} \bf \nabla \rm \theta_{5}-
                   \theta_5 \bf \nabla \rm \theta_4)\nonumber \\
      &\!\!\!\!+&\!\!\!\! \frac{e^2}{2}\bf A_{\rm T}^{\rm 2}\rm
                (\theta_{1}^{2}+\theta_{2}^{2}+
                 \theta_{4}^{2}+\theta_{5}^{2})
                 -\frac{N_{c}e^2}{6\pi^2 f_\pi}
                 [\bf E_{\rm T}\cdot B\rm +
                 (\frac{\bf \nabla \rm}{\bf \nabla \rm^2}\rm \rho^{(2)})
                  \bf\cdot B\rm](\theta_3+\frac{1}{\sqrt{3}}
                              \theta_{8}) \; \; ,
\end{eqnarray}
is the expression for the hamiltonian density of the model.
We see that without mentioning gauge fixing the solution of the
constraint lead to a Coulomb gauge hamiltonian density $.\;\;\;$
( Note that $ \bf B=\nabla\times A_{\rm T} $ ).
%
%
\section{Keeping third order terms}
\label{exp-third}

We now proceed with the expansion and keep terms up to the third
order in the Goldstone boson fields. We obtain the following
expression for the resulting $ U_{EM} (1) $ gauge invariant effective
lagrangian density
\begin{equation}
    {\cal L}_{eff}={\cal L}_{EM}+{\cal L}_{\sigma}^{(2)}+
                   {\cal L}_{WZW}^{(2)}+{\cal L}_{WZW}^{(3)}
                   +O(\theta^4)\; \; ,
\end{equation}
where the expression for ${\cal L}_{\sigma}^{(2)}$ and
${\cal L}_{WZW}^{(2)}$ are given in (3) and
\begin{eqnarray}
   {\cal L}_{WZW}^{(3)}&\!\!\!\!=&\!\!\!\!
               -\frac{N_{c}e}{3\pi^{2}f_{\pi}^3}
               {\epsilon^{\mu \nu \alpha \beta}}(\partial_\mu A_\nu)
              (\theta_1\partial_\alpha \theta_2 -
                \theta_2 \partial_\alpha \theta_1+
                \theta_4\partial_\alpha \theta_5 -
                \theta_5 \partial_\alpha \theta_4)
               (\partial_\beta \theta_3+\frac{1}{\sqrt{3}}
                 \partial_\beta \theta_8) \nonumber \\
   &\!\!\!\!-&\!\!\!\! \frac{N_{c}e}{\sqrt{3}\pi^{2}f_{\pi}^3}
               {\epsilon^{\mu \nu \alpha \beta}}(\partial_\mu A_\nu)
               (\theta_7\partial_\alpha \theta_6 -
                \theta_6 \partial_\alpha \theta_7)\partial_\beta
                \theta_8 \nonumber \\
        &\!\!\!\!+&\!\!\!\! \frac{N_{c}e^2}{18\pi^{2}f_{\pi}^3}
                {\epsilon^{\mu \nu \alpha \beta}}(\partial_\mu A_\nu)
                      (\partial_\alpha A_\beta)
               \{ [4(\theta_{1}^2+\theta_{2}^2)+
                5(\theta_{4}^2+\theta_{5}^2)] \theta_3 \nonumber \\
        &\!\!\!\!+&\!\!\!\!
             \sqrt{3}[2(\theta_{1}^2+\theta_{2}^2)+
                \theta_{4}^2+\theta_{5}^2]\theta_8 +
         2[(\theta_1 \theta_5 - \theta_2 \theta_4)\theta_7 +
          (\theta_1 \theta_4 + \theta_2 \theta_5)\theta_6] \}
                    \nonumber \\
        &\!\!\!\!-&\!\!\!\! \frac{N_{c}e^2}{3\pi^{2}f_{\pi}^3}
                {\epsilon^{\mu \nu \alpha \beta}}
                A_\mu (\partial_\nu A_\alpha)
                (\theta_3 + \frac{1}{\sqrt{3}}\theta_{8})
                \partial_\beta (\theta_{1}^2+\theta_{2}^2 +
                \theta_{4}^2+\theta_{5}^2) \nonumber  \; \; .
\end{eqnarray}
We see that only $\Gamma_{WZW} (U,A_\mu) $ ( the part of the
Wess-Zumino-Witten action which describes the anomalous
interaction of the Goldstone bosons with the photons ) gives
extra contribution at this order.
 Next we derive the expressions for the canonical
momenta $p_a$ conjugate to the Goldstone boson fields
$\theta_a$. We substitute into (13) and we get the following
 expression for the effective lagrangian density in the enlarged
configuration space
\vspace{1em}
\begin{equation}
   {\cal L}_{eff}=-\bf E\cdot \dot{A}\rm +
                  p_{\it a}\dot{\theta_{\it a}}
                 - H^{(2)}-H^{(3)}-A_{0}(\rho^{(2)}+\rho^{(3)}
                 -\bf\nabla\cdot E\rm)+O(\theta^4)\; \; ,
\vspace{1em}
\end{equation}
\vspace{1em}
where the expression for $H^{(2)}$ is given in (12) and
\begin{eqnarray}
\vspace{1em}
   H^{(3)}&\!\!\!\!=&\!\!\!\!-\frac{N_{c}e}{3\pi^{2}f_{\pi}^3}
              [\bf E\times\bf \nabla \rm (\theta_3 +
              \frac{1}{\sqrt{3}} \theta_8)-
              (p_3+\frac{1}{\sqrt{3}}p_8) \bf B\rm)]
              \cdot(\theta_1\bf \nabla \rm \theta_2-\theta_2\bf \nabla
              \rm\theta_1+\theta_4\bf \nabla \rm \theta_5-
              \theta_5\bf \nabla \rm\theta_4) \nonumber \\
          &\!\!\!\!-&\!\!\!\!\frac{N_{c}e}{\sqrt{3}\pi^{2}f_{\pi}^3}
              (\bf E\times\bf \nabla \rm \theta_8-p_8 \bf B\rm)
              \cdot(\theta_7\bf \nabla \rm \theta_6-\theta_6\bf \nabla
              \rm\theta_7) \nonumber \\
      &\!\!\!\!+&\!\!\!\! \frac{N_{c}e^2}{9\pi^{2}f_{\pi}^3}
              (\bf E\cdot B\rm)
              \{ [4(\theta_{1}^2+\theta_{2}^2)+
                5(\theta_{4}^2+\theta_{5}^2)] \theta_3 \nonumber \\
        &\!\!\!\!+&\!\!\!\!
             \sqrt{3}[2(\theta_{1}^2+\theta_{2}^2)+
                \theta_{4}^2+\theta_{5}^2]\theta_8 +
         2[(\theta_1 \theta_5 - \theta_2 \theta_4)\theta_7 +
          (\theta_1 \theta_4 + \theta_2 \theta_5)\theta_6] \}
                    \nonumber \\
      &\!\!\!\!-&\!\!\!\! \frac{N_{c}e^2}{3\pi^{2}f_{\pi}^3}
              (\bf E\times A\rm) \cdot [\bf \nabla
          \rm(\theta_{1}^2+\theta_{2}^2+\theta_{4}^2+\theta_{5}^2)]
             (\theta_3+\frac{1}{\sqrt{3}}\theta_8)\nonumber \\
      &\!\!\!\!-&\!\!\!\! \frac{2N_{c}e^2}{3\pi^{2}f_{\pi}^3}
            (\bf A\cdot B\rm)(p_1\theta_1+p_2\theta_2+p_4\theta_4+
            p_5\theta_5)(\theta_3+\frac{1}{\sqrt{3}}\theta_8)\nonumber \\
       &\!\!\!\!-&\!\!\!\! \frac{N_{c}e}{3\pi^{2}f_{\pi}^3}
           [\bf B\cdot\bf \nabla \rm(\theta_3+\frac{1}{\sqrt{3}}\theta_8)]
           (p_2\theta_1-p_1\theta_2+p_5\theta_4-p_4\theta_5)\nonumber \\
       &\!\!\!\!-&\!\!\!\! \frac{N_{c}e}{\sqrt{3}\pi^{2}f_{\pi}^3}
           (\bf B\cdot\bf \nabla \rm \theta_8)
           (p_6\theta_7-p_7\theta_6) \; \; , \\
\nonumber \\
\nonumber \\
  \rho^{(2)}&\!\!\!\!=&\!\!\!\! e(p_2\theta_1-p_1\theta_2+
             p_5\theta_4-p_4\theta_5)\; \; ,\nonumber \\
\nonumber \\
  \rho^{(3)}&\!\!\!\!=&\!\!\!\!-\frac{N_{c}e^2}{3\pi^{2}f_{\pi}^3}
                      \bf \nabla \rm\cdot[\bf B\rm
              (\theta_{1}^2+\theta_{2}^2+\theta_{4}^2+\theta_{5}^2)
              (\theta_3+\frac{1}{\sqrt{3}}\theta_8)] \; \; .\nonumber
\end{eqnarray}
The scalar potential $A_{0}$ is again the Lagrange multiplier
and the equation of the constraint is given by
\begin{equation}
   \bf \nabla\cdot E\rm-(\rho^{(2)}+\rho^{(3)})=0\; \; .
\end{equation}
The equation (16) has similar structure as in the previous case (7)
so we proceed similarly. We decompose the electric field $\bf E$
and the vector potential
$\bf A$ into transverse and longitudinal components. Then (16)
implies
\begin{equation}
   \bf E_{\rm L}=\rm \frac{\bf \nabla \rm}{\bf \nabla
                 \rm^2}(\rho^{(2)}+\rho^{(3)})
                \equiv-\frac{1}{4\pi}\int\!d^3\!y
\frac{1}{\mid \bf y \rm -\bf x \rm \mid}\nabla
(\rho^{(2)}(y) + \rho^{(3)}(y)) \; \; .
\end{equation}
Next we substitute the expression for $\bf E_{\rm L}$ given in
(17) into (14) and we obtain a lagrangian density whose canonical
one-form is given (apart from a total divergence) by
\begin{equation}
  -\bf E_{\rm T}\cdot \dot{A}_{\rm T}\rm
  +\rho^{(2)}\frac{\bf \nabla \rm}{\bf \nabla \rm^2}
                   \cdot\dot{\bf A}_{\rm L}\rm
  +\rho^{(3)}\frac{\bf \nabla \rm}{\bf \nabla \rm^2}\cdot
                   \dot{\bf A}_{\rm L}\rm
  +p_{\it a} \dot{\theta}_{\it a}\; \; ,
\end{equation}
This expression must be diagonalized so that it acquires the
standard form (6). In order to do this we proceed in two steps
as in the SU(2) case \cite{PP}.
First we perform the Darboux's transformations given in (10)
which lead to partial diagonalization of (18). The new lagrangian
density has a canonical one-form given by
\begin{displaymath}
  -\bf E_{\rm T}\cdot \dot{A}_{\rm T}\rm
  +\rho^{(3)}\frac{\bf \nabla \rm}{\bf \nabla \rm^2}
                  \cdot\dot{\bf A}_{\rm L}\rm
  +p_{\it a} \dot{\theta}_{\it a}\; \; ,
\end{displaymath}
The term
$\rho^{(3)}\frac{\bf \nabla \rm}{\bf \nabla \rm^2}
\cdot\dot{\bf A}_{\rm L}\rm$ can be written as follows apart from
a total time derivative
\begin{displaymath}
  \rho^{(3)}\frac{\bf \nabla \rm}{\bf \nabla \rm^2}
                  \cdot\dot{\bf A}_{\rm L}\rm =
                 - \frac{N_{c}e^2}{3\pi^2 f_{\pi}^3}
     [(\bf \nabla \rm{\phi}\times\bf A_{\rm L})\cdot\dot{A}_{\rm T}
       \rm +(\bf B\cdot A_{\rm L}\rm)\dot{\phi}]\; \; ,
\end{displaymath}
where $\phi=(\theta_{1}^2+\theta_{2}^2+\theta_{4}^2+\theta_{5}^2)
(\theta_3+\frac{1}{\sqrt{3}}\theta_8)\; \; .$
\vspace{1em}

Now we proceed with the second step. We perform the following
Darboux's transformations
\vspace{1em}
\begin{displaymath}
 \begin{array}{l}
\vspace{0.5em}
  \bf E_{\rm T}\rightarrow E_{\rm T}-
          \rm \frac{N_{c}e^2}{3\pi^{2}f_{\pi}^3}
          \bf \nabla \rm[(\theta_{1}^2+\theta_{2}^2
          +\theta_{4}^2+\theta_{5}^2)(\theta_3+
           \frac{1}{\sqrt{3}}\theta_8)]\bf\times A_{\rm L}\;\;,\\
\vspace{0.5em}
  p_i \rightarrow p_i +\frac{2N_{c}e^2}{3\pi^{2}f_{\pi}^3}
               (\bf B\cdot A_{\rm L}\rm)
(\theta_3+\frac{1}{\sqrt{3}}\theta_8)\theta_i \;\;,\;\;
                                            i=1,2,4,5 \\
\vspace{0.5em}
  p_3 \rightarrow p_3 +\frac{N_{c}e^2}{3\pi^{2}f_{\pi}^3}
               (\bf B\cdot A_{\rm L}\rm)(\theta_{1}^2+
          \theta_{2}^2+\theta_{4}^2+\theta_{5}^2)\;\;,\\
\vspace{0.5em}
  p_8 \rightarrow p_8 +\frac{N_{c}e^2}{3\sqrt{3}\pi^{2}f_{\pi}^3}
               (\bf B\cdot A_{\rm L}\rm)(\theta_{1}^2+
          \theta_{2}^2+\theta_{4}^2+\theta_{5}^2)\;\;,\\
\vspace{0.5em}
     \bf A_{\rm T} \rightarrow A_{\rm T}\rm \;\; ,\;\;
     p_6\rightarrow p_6 \;\; , \;\;
     p_7\rightarrow p_7 \;\; , \;\;
     \theta_{\it a}\rightarrow\theta_{\it a}
                    \; \;,\;\;{\it a}=1,...,8 \;\; .
 \end{array}
\end{displaymath}
These transformations complete the diagonalization of the
canonical one-form and we end up with a lagrangian density
where the longitudinal part of
the vector potential $\bf A_{\rm L}$ cancels out exactly
\vspace{1em}
\begin{equation}
  {\cal L}=-\bf E_{\rm T}\cdot \dot{A}_{\rm T}\rm +
           p_{\it a}\dot{\theta_{\it a}}
          - H_{C}^{(2)}-H_{C}^{(3)}+O(\theta^4)\; \; .
\vspace{1em}
\end{equation}
The expression for $H_{C}^{(2)}$ is given in (12) and
\begin{eqnarray}
   H_{C}^{(3)}&\!\!\!\!=&\!\!\!\!-\frac{N_{c}e}{3\pi^{2}f_{\pi}^3}
              [\bf E_{\rm T}\times\bf \nabla \rm (\theta_3 +
              \frac{1}{\sqrt{3}} \theta_8)-
              (p_3+\frac{1}{\sqrt{3}}p_8) \bf B\rm)]
              \cdot(\theta_1\bf \nabla \rm \theta_2-\theta_2\bf \nabla
              \rm\theta_1+\theta_4\bf \nabla \rm \theta_5-
              \theta_5\bf \nabla \rm\theta_4) \nonumber \\
      &\!\!\!\!-&\!\!\!\!\frac{N_{c}e}{\sqrt{3}\pi^{2}f_{\pi}^3}
            (\bf E_{\rm T}\times\bf \nabla \rm \theta_8-p_8 \bf B\rm)
            \cdot(\theta_7\bf \nabla \rm \theta_6-\theta_6\bf \nabla
            \rm\theta_7) \nonumber \\
      &\!\!\!\!+&\!\!\!\! \frac{N_{c}e^2}{9\pi^{2}f_{\pi}^3}
              (\bf E_{\rm T}\cdot B\rm)
              \{ [4(\theta_{1}^2+\theta_{2}^2)+
                5(\theta_{4}^2+\theta_{5}^2)] \theta_3 \nonumber \\
      &\!\!\!\!+&\!\!\!\!
             \sqrt{3}[2(\theta_{1}^2+\theta_{2}^2)+
             \theta_{4}^2+\theta_{5}^2]\theta_8 +
             2[(\theta_1 \theta_5 - \theta_2 \theta_4)\theta_7 +
             (\theta_1 \theta_4 + \theta_2 \theta_5)\theta_6] \}
                    \nonumber \\
      &\!\!\!\!-&\!\!\!\! \frac{N_{c}e^2}{3\pi^{2}f_{\pi}^3}
          (\bf E_{\rm T}\times A_{\rm T}\rm) \cdot [\bf \nabla
          \rm(\theta_{1}^2+\theta_{2}^2+\theta_{4}^2+\theta_{5}^2)]
          (\theta_3+\frac{1}{\sqrt{3}}\theta_8)\nonumber \\
      &\!\!\!\!-&\!\!\!\! \frac{2N_{c}e^2}{3\pi^{2}f_{\pi}^3}
         (\bf A_{\rm T}\cdot B\rm)(p_1\theta_1+p_2\theta_2+p_4\theta_4+
         p_5\theta_5)(\theta_3+\frac{1}{\sqrt{3}}\theta_8)\nonumber \\
      &\!\!\!\!-&\!\!\!\! \frac{N_{c}e}{3\pi^{2}f_{\pi}^3}
         [\bf B\cdot\bf \nabla \rm(\theta_3+\frac{1}{\sqrt{3}}\theta_8)]
         (p_2\theta_1-p_1\theta_2+p_5\theta_4-p_4\theta_5)\nonumber \\
      &\!\!\!\!-&\!\!\!\! \frac{N_{c}e}{\sqrt{3}\pi^{2}f_{\pi}^3}
           (\bf B\cdot\bf \nabla \rm \theta_8)
           (p_6\theta_7-p_7\theta_6) \; \; .\\
\nonumber
\end{eqnarray}
$ H_{C}^{(2)}+H_{C}^{(3)}$ is the expression for the
Coulomb-gauge hamiltonian density. In terms of the usual
pseudoscalar fields $\pi^\pm =\frac{1}{\sqrt{2}}(\theta_1 \mp
i\theta_2)\;,\;\; K^\pm =\frac{1}{\sqrt{2}}(\theta_4 \mp
i\theta_5)\;,\;  K^\circ =\frac{1}{\sqrt{2}}(\theta_6 -
i\theta_7)\;,\; \overline{K}^\circ =\frac{1}{\sqrt{2}}(\theta_6 +
i\theta_7)\;\;,\;\;
\pi^\circ=\theta_3\;\;,\;\;\eta_8=\theta_8\;\;$ and the
corresponding canonical momenta $p_i\;\;(i=\pi^\pm\;,\;
\pi^\circ\;,\;K^\pm\;,\;\overline{K}^\circ\;,\;K^\circ\;,\;\eta_8)\;\;$
the expressions for $H_{C}^{(2)}$ and $H_{C}^{(3)}$ are given by
\begin{eqnarray}
   H_{C}^{(2)}&\!\!\!\!=&\!\!\!\!\frac{1}{2}
        [\bf E_{\rm T}^{\rm 2}+B^{\rm 2}\rm-
                  \rho^{(2)}\frac{1}{\bf \nabla \rm^2}
                  \rho^{(2)}
                  +p_{\pi^ \circ}^{2}+p_{\eta_8}^{2}
+2p_{\pi^{+}} p_{\pi^{-}}+2p_{K^{+}} p_{K^{-}}+
 2p_{K^\circ}p_{\overline{K}^\circ}\nonumber \\
       &\!\!\!\!+&\!\!\!\!(\bf \nabla \rm\pi^{\circ})^{2}
+(\bf \nabla \rm\eta_8)^{2}+
2(\bf\nabla\rm\pi^{+})\cdot(\bf\nabla\rm\pi^{-})+
2(\bf\nabla\rm K^{+})\cdot(\bf\nabla\rm K^{-})+
2(\bf\nabla\rm K^\circ)\cdot(\bf\nabla\rm \overline{K}^\circ)]\nonumber \\
          &\!\!\!\!-&\!\!\!\! ie\bf A_{\rm T}\cdot\rm
                 (\pi^{+}\nabla\pi^{-}- \pi^{-}\nabla\pi^{+}
                 + K^{+}\nabla K^{-}- K^{-}\nabla K^{+} )
                 + e^2 \bf A_{\rm T}^{\rm 2}\rm
                    (\pi^{+}\pi^{-} + K^{+}K^{-}) \nonumber \\
          &\!\!\!\!-&\!\!\!\!\frac{N_{c}e^2}{6\pi^2 f_\pi}
                [\bf E_{\rm T}\cdot B\rm +
                (\frac{\bf \nabla \rm}{\bf \nabla \rm^2}\rm \rho^{(2)})
                 \bf\cdot B\rm](\pi^\circ+\frac{1}{\sqrt{3}}
                              \eta_8) \; \;,\nonumber \\
\nonumber \\
\nonumber \\
   H_{C}^{(3)}&\!\!\!\!=&\!\!\!\!\frac{iN_{c}e}{3\pi^{2}f_{\pi}^3}
              [\bf E_{\rm T}\times\bf \nabla \rm (\pi^\circ +
              \frac{1}{\sqrt{3}} \eta_8)-
              (p_{\pi^\circ}+\frac{1}{\sqrt{3}}p_{\eta_8}) \bf B\rm)]
              \cdot(\pi^{+}\nabla \pi^{-}+K^{+}\nabla K^{-}) \nonumber \\
       &\!\!\!\!+&\!\!\!\!\frac{iN_{c}e}{\sqrt{3}\pi^{2}f_{\pi}^3}
           (\bf E_{\rm T}\times\bf \nabla \rm \eta_8-p_{\eta_8} \bf B\rm)
\cdot(\overline{K}^\circ \nabla \overline{K}^\circ) \nonumber \\
      &\!\!\!\!+&\!\!\!\! \frac{N_{c}e^2}{9\pi^{2}f_{\pi}^3}
              (\bf E_{\rm T}\cdot B\rm)
              [(4\pi^{+}\pi^{-}+5K^{+}K^{-}) \pi^\circ
             + \sqrt{3}(2\pi^{+}\pi^{-}+K^{+}K^{-})\eta_8+
              2\sqrt{2}\pi^{+}K^{-}K^\circ] \nonumber \\
      &\!\!\!\!-&\!\!\!\! \frac{N_{c}e^2}{3\pi^{2}f_{\pi}^3}
              (\bf E_{\rm T}\times A_{\rm T}\rm) \cdot [\bf \nabla
               \rm(\pi^{+}\pi^{-}+K^{+}K^{-})]
             (\pi^\circ+\frac{1}{\sqrt{3}}\eta_8)\nonumber \\
      &\!\!\!\!-&\!\!\!\! \frac{2N_{c}e^2}{3\pi^{2}f_{\pi}^3}
            (\bf A_{\rm T}\cdot B\rm)(p_{\pi^{+}}\pi^{+} +
             p_{K^{+}}K^{+})(\pi^\circ+\frac{1}{\sqrt{3}}\eta_8)
                                  \nonumber \\
      &\!\!\!\!+&\!\!\!\! \frac{iN_{c}e}{3\pi^{2}f_{\pi}^3}
           [\bf B\cdot\bf \nabla \rm(\pi^\circ+\frac{1}{\sqrt{3}}\eta_8)]
           (p_{\pi^{+}}\pi^{+}+p_{K^{+}}K^{+})\nonumber \\
      &\!\!\!\!+&\!\!\!\! \frac{iN_{c}e}{\sqrt{3}\pi^{2}f_{\pi}^3}
           (\bf B\cdot\bf \nabla \rm \eta_8)
           p_{\overline{K}^\circ}\overline{K}^\circ
           \;\;\;\; + \;\;\;\; h.c. \; \; ,\nonumber \\
\nonumber \\
\nonumber \\
           \rho^{(2)}&\!\!\!\!=&\!\!\!\! -ie(p_{\pi^{+}}\pi^{+}-
           p_{\pi^{-}}\pi^{-}+p_{K^{+}}K^{+}-p_{K^{-}}K^{-})
                       \;\;.\\
\nonumber
\end{eqnarray}
%
%
\section{Conclusion}
\label{Conc}

In this work we applied the Faddeev and Jackiw method for
constrained systems to the $U_{EM}(1)$ gauged three flavour WZW
model. We expanded the U-field
into series of powers of the Goldstone boson fields and we
limited ourselves to lagrangian densites with terms up to the
third power in $\theta_{\it a}$. After treating these constrained
langrangian densities using the FJ
approach we ended up with unconstrained Coulomb-gauge hamiltonians,
as in the case of QED \cite{F-J}. The general case which
involves any number of Goldstone boson fields is currently under
investigation.
%
%
\section{Appendix}
\label{app}

Our metric is $g_{\mu \nu}=diag(1,-1,-1,-1) \;\; , \;\;
Q=diag(2/3,-1/3,-1/3)$
is the charge matrix,
 $D_\mu=\partial_\mu + ieA_\mu [Q,\;\;]$
denote the covariant derivative.
 $\lambda_{a}\hspace{0.25em},\hspace{0.25em} a=1,...,8$ are the
SU(3) generators $ \mbox{tr}\,(\lambda_a \lambda_b)=2\delta_{ab}$.
 We choose $ e>0 $ so that the electric charge of
the electron is $-e$. We define $ \epsilon^{0123}=1$ .

\end{document}